\documentclass[journal]{IEEEtran}

\usepackage[english]{babel}
\usepackage[utf8]{inputenc}
\usepackage[T1]{fontenc}
\usepackage[pdftex]{graphicx}
\usepackage{color}
\usepackage{supertabular}

\ifCLASSINFOpdf
\else
\fi

\begin{document}

\title{Towards Understanding the Impact of Human Mobility on Police Allocation}

\author{Carlos Caminha, Vasco Furtado}

\markboth{}%
{}

\maketitle

\begin{abstract}

Motivated by recent findings that human mobility is proxy for crime behavior in
big cities and that there is a superlinear relationship between the people’s
movement and crime, this article aims to evaluate the impact of how these
findings influence police allocation. More precisely, we shed light on the
differences between an allocation strategy, in which the resources are
distributed by clusters of floating population, and conventional allocation
strategies, in which the police resources are distributed by an Administrative
Area (typically based on resident population). We observed a substantial
difference in the distributions of police resources allocated following these
strategies, what evidences the imprecision of conventional police allocation
methods. 

\end{abstract}


\IEEEpeerreviewmaketitle

\section{Introduction}

Faced with the ever-growing problem of crime, prevention strategies have come to
the fore as a key issue and one of the main challenges of Law Enforcement
authorities. This increase has been observed mainly in large urban centers and
even with huge masses of digitized data about police activities and reported
crimes, the police institutions do not seem to use adequately this information
to fight the growth of crime.

Within this context, strategies of police allocation play an important role in
crime prevention and a recent discovery by Caminha {\it et al.}
\cite{caminha2017} motivated us to revisit the state of the art of this subject.
Caminha {\it et al.} discovered a superlinear relationship
\cite{bettencourt2010,gomez2012,ignazzi2014,hanley2016,alves2015a,arcaute2015,
arcaute2016,cottineau2016,leitao2016,bettencourt2013,arbesman2009} between the
flow of people and property crimes. In other words, the authors found that the
increase of the floating population in a urban space implies a disproportionate
growth of property crime in that space. Formally this relationship can be
represented by the equation $Y=aX^\beta$, indicating a Power Law, where $Y$
quantifies property crimes, $X$ quantifies the volume of people flow, $a$ is a
normalization constant and $\beta$ is the exponent that scales the relation,
which in the case of a superlinear relation is assumed to be $\beta > 1$. The
authors further assert that administrative territorial units that typically
account for features of resident population, such as divisions by
neighborhoods, census tracts or zones, are unable to precisely capture the
effect that social relations have over crime. This fact has already been stated
by several urban indicators in important scientific productions \cite{makse1998,
rozenfeld2008, giesen2010, rozenfeld2011, duranton2013, gallos2012,
duranton2015, eeckhout2004, oliveira2014}.

Although over the years a series of scientific works have studied factors
surrounding crime \cite{melo2014,gelman2007,agnew2007,caminha2012,
furtado2012,guedes2014,kennedy1990,beato2014}, none of them have taken into
consideration this finding that quantifies the relation between human mobility
and crime. More specifically, they do not make use of the divisions of urban
spaces estimated from the floating population to police allocation.

In this article we seek to understand the impact, in the allocation of police
resources, from the fact that the relationship between the movement of people
and property crimes follows a Power Law. To estimate this impact, we use data
from a big metropolis to build clusters of floating population that will be
considered as the basis for the allocation strategy. The distribution of the
police resources obtained from the application of this strategy is compared to a
conventional allocation strategy, in which police officers are distributed into
administrative territorial divisions. Doing so, we were able to show the
difference between the distribution of allocated police according to the two
strategies. This difference allows us to conclude that, under the light of these
new evidences of cause-effect between floating population and property crimes,
it is inaccurate to apply a conventional strategy of police allocation, which is
based only on resident population.

\section{State of Art}

There is a vast selection of literature on police allocation in urban space to
combat criminal activity. There was a growing interest in developing techniques
using programs of spatial analysis to identify areas where the police resources
are to be allocated. In a very general way, a typical strategy of allocation is
to implement a heterogeneous model, in which the distribution of resources in a
geographic area is directly proportional to the density of crimes of that
region. Typically these areas are administrative regions ({ \it e.g.} census
tract or neighbohoods) demarcated from features of the resident population
\cite{sherman1989}. This perspective, is not totally in line with routine
activity theory \cite{clarke1993,michael1933,cohen1979} and criminal career
approaches \cite{blumstein1986},but ,for practical reasons, have been used for
years \cite{sherman1995}.

With the increase in the volume of digital data and the creation of more
sophisticated mapping techniques, opportunities have appeared to go beyond the
approaches where only the density of crimes in areas of resident population is
considered \cite{weisburd2006,ratcliffe2006,wortley2016}. Nevertheless, much of
the work in this area continued to focus on the concentration of crime in
administrative territorial units \cite{groff2002,berk2011}. It is true that
Kennedy {\it et al.} \cite{kennedy2011} developed an in-depth assessment of
social factors that contribute to crime occurrence. However, its allocation
algorithm is based on risk areas which indirectly are also measured by resident
population indicators.

There are also a number of papers that use simulation models to teach the police
officer how to make an allocation of quality resources
\cite{greasley1998,melo2005,furtado2006,reis2006,guedes2015}. However, these
works do not consider the new evidence that human mobility is the key to
understanding the emergence of property crimes in regions of urban space.

Finally, it is worth noting that there are numerous studies that seek to
understand phenomena related to human mobility
\cite{gonzalez2008,wang2011,caminha2016,andrade2009,ponte2016}, however, works
that apply the knowledge obtained from these studies on crime prevention from
police allocation is scarce.

\section{Datasets}
In this paper, data on property crimes was used, this was obtained from
\cite{ciops2016}. In total this dataset contains 81,911 geo-referenced crimes
occurring between August 2005 and July 2007. Three levels of segmentation were
used for the city of Fortaleza-CE, Brazil. The first level was a division by
neighborhood, obtained from \cite{bairros2017}, in total, Fortaleza has 116
districts spread over an area of 313 $\mathrm{km^2}$ where more than 2,400,000
people live. The second level, a division by defined census tracts by IBGE
(Brazilian Institute of Geography and Statistics) \cite{ibge2016}, which divides
the city into 3043 subareas that on average contain 800 residents each. Finally,
the third level of segmentation, a division by clusters of floating population,
estimated in \cite{caminha2017}. In total, the authors divided Fortaleza into
119 clusters using \textit{City Clustering Algorithm} (CCA) \cite{makse1998,
rozenfeld2008, giesen2010, rozenfeld2011, duranton2013, gallos2012,
duranton2015, eeckhout2004}.

\begin{figure}[!h]
\includegraphics[width=0.486\textwidth]{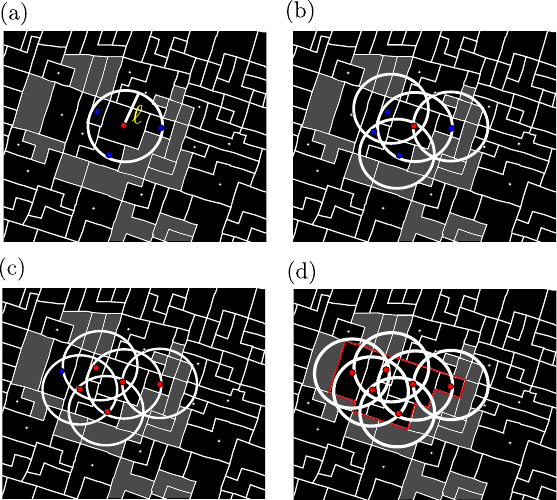}
\caption{{\bf The scheme of the City Clustering Algorithm (CCA).} Each square
represents a clustering unit, specifically in our case, they represent census
tracts. Black squares are candidates for clustering $(D_i > D^*)$, in contrast,
the gray squares cannot be clustered ($D_i \le D^*$). (a) The red dot represents
the geometric center of the $i$-th census tract and the white circle with radius
$\ell$ seeks neighbors belonging to the same cluster. (b) The same search
operation is made for the other census tracts. (c) The same operation is done
until there are no more neighbors within the radius of operation. (d) The
algorithm finishes running and the cluster is found.}
\label{fig1}
\end{figure}

To define the boundaries of this clusters, the CCA algorithm considered the
notion of spatial continuity through the aggregation of census tracts that are
near one another. The CCA constructs the floating population boundaries of an
urban area considering two parameters, namely, a population density threshold,
$D^*$, and a distance threshold, $\ell$. For the $i\mathrm{-th}$ census tract,
the population density $D_i$ is located in its geometric center; if $D_i > D^*$,
then the $i\mathrm{-th}$ census tract is considered populated. The length $\ell$
represents a cutoff distance between census tracts to consider them as spatially
contiguous, {\it i.e.}, all of the nearest neighboring census tracts that are at
distances smaller than $\ell$ are clustered. Hence, a cluster made by the CCA is
defined by populated areas within a distance less than $\ell$, as seen
schematically in Figure \ref{fig1}. Previous studies
\cite{oliveira2014,duranton2013, duranton2015} have demonstrated that the
results produced by the CCA can be weakly dependent on $D^*$ and $\ell$ for some
range of parameter values. In \cite{caminha2017} $\ell$ was quantified in meters
(m) and $D^*$ in people passing by $\mathrm{km^2}$ in one day.

\begin{figure}[!h]
\includegraphics[width=0.48\textwidth]{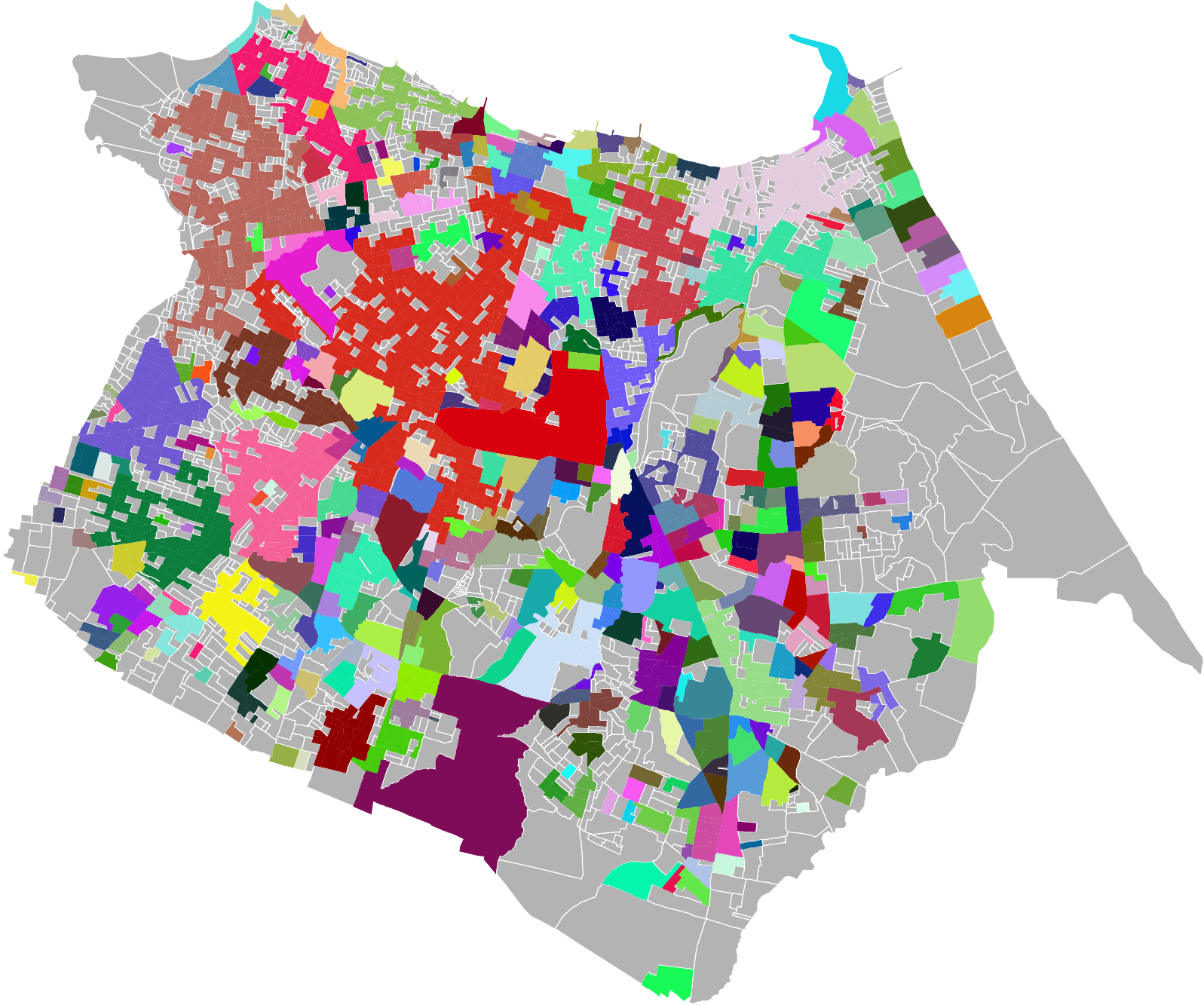}
\caption{{\bf Agglomerates found by the CCA algorithm.} The regions in light
gray color were not grouped because they have low floating population density
($D_i \le D^*$), the other colors represent clusters found. More precisely, each
color represents a people flow agglomerate.} 
\label{fig2}
\end{figure}

Figure \ref{fig2} illustrates the clusters found. The base division used in the
cluster was the census tract map. The census tract in light gray color were not
grouped because they have low flux density ($D_i \le D^*$), the other colors
represent clusters found. In the division reached by the CCA the volume of flow
of a cluster is proportional to its area \cite{oliveira2014}. It was estimated
$\ell=320$ and $D^*=6000$.

\section{Methods}

Two strategies of police allocation will be compared here, these strategies are
based on the most popular heterogeneous allocation model, namely by high crime
density. The first, called \textit{Resident Population Allocation (RPA)
Strategy} is a conventional strategy of police allocation, whose resources are
distributed in proportion to the quantity of occurrences in administrative
divisions of a territory (what is typically estimated from features from the
resident population). In this work the division by neighborhood's boundaries
will be adopted, because, despite the division by census tracts being available,
it is too segmented, with some of them being less than one block, thus being
unfeasible to be used in a real policy of resource allocation.

The second allocation strategy, called \textit{Floating Population Allocation
(FPA) Strategy}, will also distribute police resources proportionally to the
number of calls to the police in a spatial division, however, in this strategy
the boundaries of the areas follow the clusters of floating population estimated
in \cite{caminha2017}.

In this way, the part of a police resource, $T_{s_i}$, allocated to a sub-region
of urban space (whether a clusters of floating population or a neighborhoods),
$s_i$ $\in$ $S$, from the quantity of crimes occurring in $s_i$, $C{s_i}$, can
be formally defined as $T_{s_i} = \frac{T * C{s_i}}{C}$. Where $T$ is the total
number of police officers available for allocation and $C$ is the total number
of crimes that have occurred in all the urban space available for allocation.

A policy of internal allocation was also adopted, precisely at the level of
$s_i$. Each cluster of floating population or neighborhood is composed of census
tracts and internally there is also a allocating of resources in a manner
proportional to the number of crimes of each census tract within $s_i$. In other
words, within each sub-region $s_i$, sectors with more crimes receive more
police officers. This sub-allocation policy is justified by the need to compare
the two strategies, which will be discussed later on.

\section{Results}

\begin{figure}[!h] 
\includegraphics[width=0.48\textwidth]{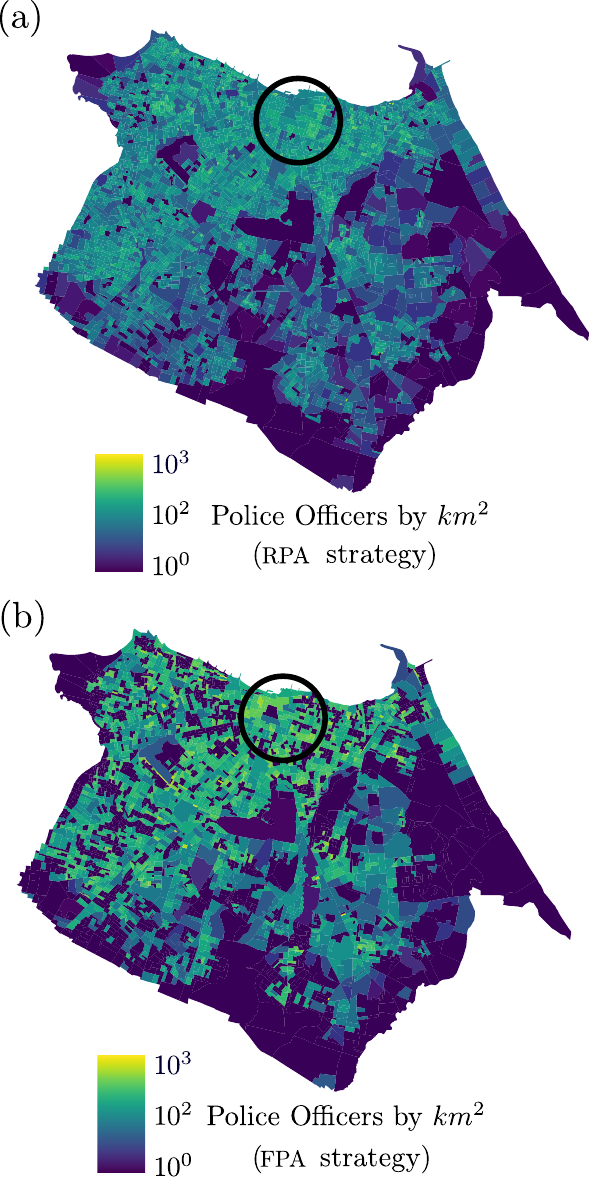}
\caption{{\bf Police allocation using the two strategies studied.} (a) shows
density map of police allocated using \textit{RPA Strategy}, (b) shows density
map of police allocated using \textit{FPA Strategy}. Black circles highlight the
shopping center of Fortaleza, an area with a large population concentration and
consequently, a large concentration of crimes against property.} 
\label{fig3}
\end{figure}

When applying \textit{RPA Strategy} and \textit{FPA Strategy} in Fortaleza to
simulate the availability of a total police resource $T=10,000$, the heat maps
shown in Figure \ref{fig3}, items (a) and (b), respectively. Hot Spots with more
intensity can be seen in \textit{FPA Strategy}, mainly in the commercial center
of the city, highlighted by the black circle in both figures. This is because
\textit{FPA Strategy} does not allocate police resources in areas that are
considered uninhabited $(D_i > D^*)$, instead concentrating more police in the
most critical regions of the city.

\begin{figure}[!h] 
\includegraphics[width=0.48\textwidth]{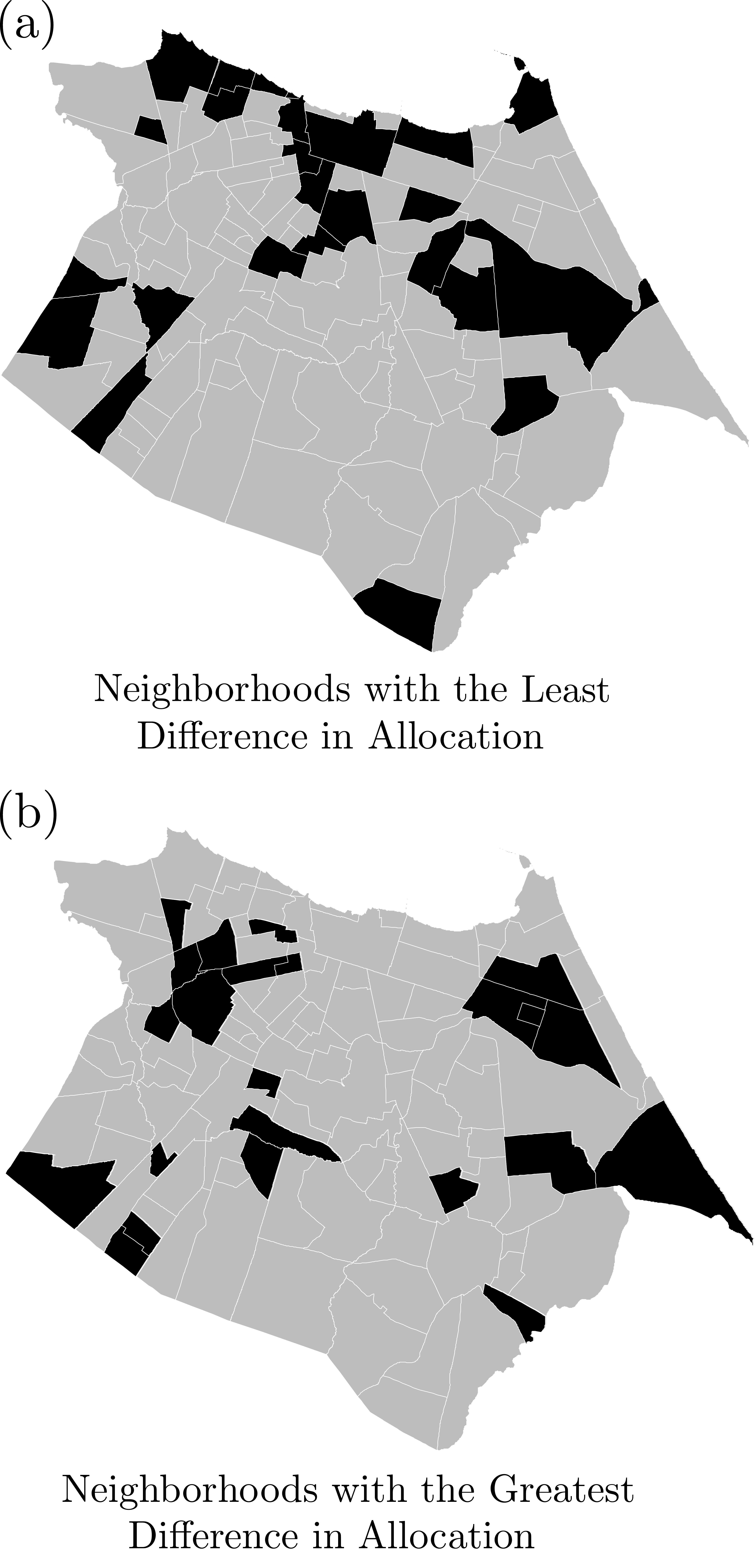}
\caption{{\bf Differences and similarities among the studied allocations.} (a)
highlights in black the neighborhoods that had the most similar allocation using
\textit{RPA e FPA Strategy}. (b) highlights the neighborhoods with the highest
difference in the number of police officers allocated. In both figures 24
neighborhoods are highlighted, 20\% of the city total.}
\label{fig4}
\end{figure}

For the purpose of comparison, the amount of police allocated per neighborhood
was calculated using \textit{FPA Strategy}. Then, the number of police officers
in the census tracts located within each neighborhood was added. After this, we
calculated the percentage difference of the number of policemen allocated by
neighborhood by both strategies. In Figure \ref{fig4}, items (a) and (b)
illustrate the neighborhoods where the allocation is more similar and more
different respectively.

In general, a greater similarity was observed in the allocations in the
neighborhoods with greater presence of fluctuating populations, these
neighborhoods are close to the commercial center of the city or located in
regions with a high concentration of residents (normally locations that are the
source of floating population). It was also observed that the districts that
presented a greater percentage difference between the quantities of police
officers allocated using the allocation strategies studied, are those which have
more non-populated census tracts, that is, with a floating population density
below the threshold $D^*$, as estimated in \cite{caminha2017}.

\begin{figure}[!h] 
\includegraphics[width=0.48\textwidth]{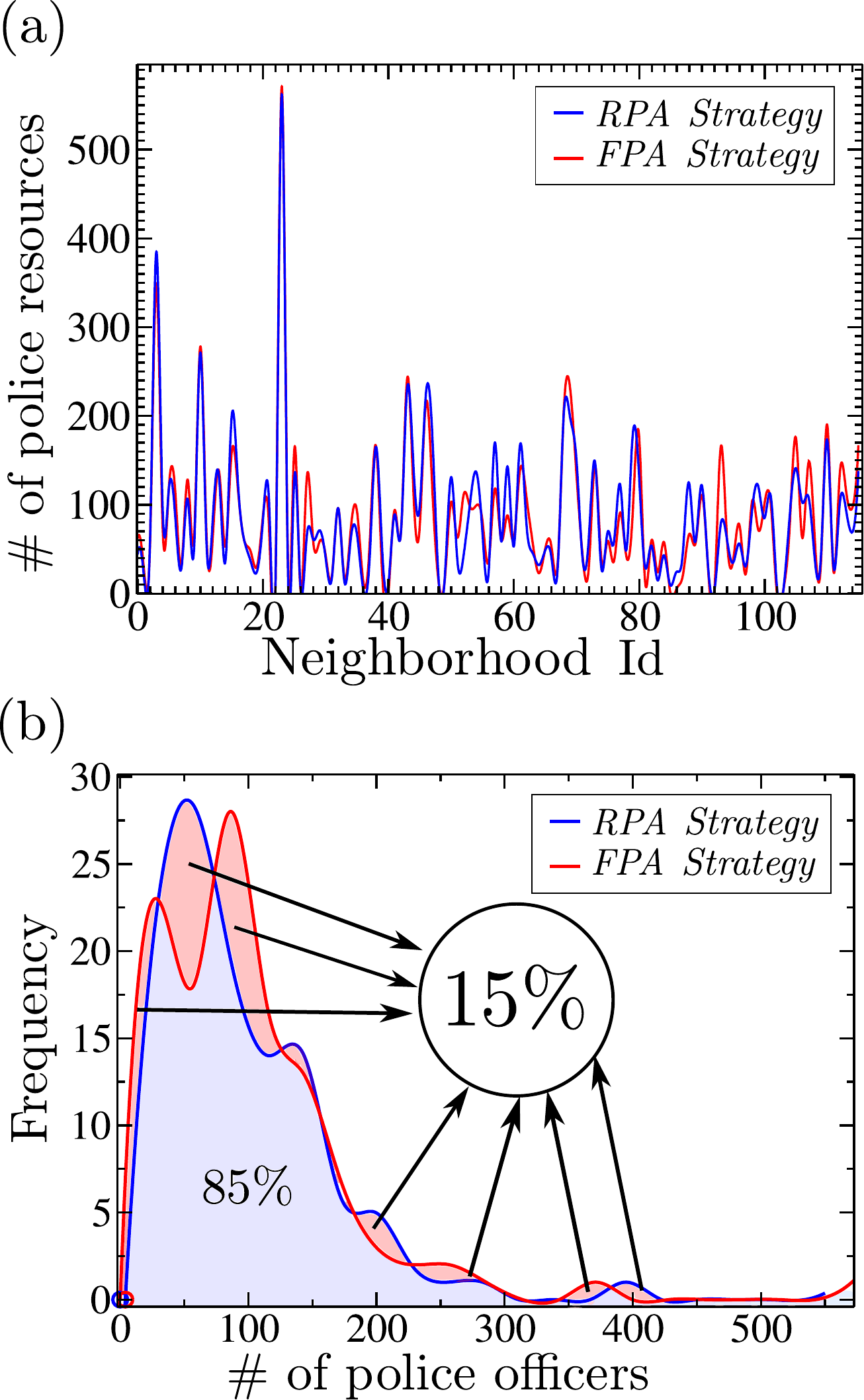}
\caption{{\bf \textit{RPA and FPA Strategy} statistical comparison.} In (a) is
illustrated the number of police resources allocated by neighborhood. The blue
line represents a \textit{Cubic Spline Interpolation} \cite{deboor1978} applied
to the values that was found in \textit{RPA Strategy}. The red line is the same
interpolation applied to the \textit{FPA Strategy}. (b) show the histograms
distribution to the allocations in the neighborhoods of the city. For better
visualization, the histograms has been generated in 20 bins \cite{wand1997}.}
\label{fig5}
\end{figure}

In Figure \ref{fig5} a more detailed comparison can be observed between the two
allocation strategies. (a) illustrates the interpolation functions
\cite{deboor1978} of the neighborhoods by the number of police officers
allocated by the two strategies studied. The intersection of the areas formed by
interpolation curves and the $x$ axis reveals approximately 15\% dissimilarity
between the allocations. This dissimilarity can be observed more clearly in
Figure \ref{fig5} (b), where the interpolation functions of the histograms
generated from the number of police officers allocated by neighborhoods
according to the two strategies is shown. The blue line represents the
interpolation function of the \textit{RPA Strategy} data. The red line
represents the estimated function for the \textit{FPA Strategy}. The regions in
light red color represent areas where there was no intersection. Added
together, these regions represent 15\% of the total area.

\begin{figure}[!h] \includegraphics[width=0.48\textwidth]{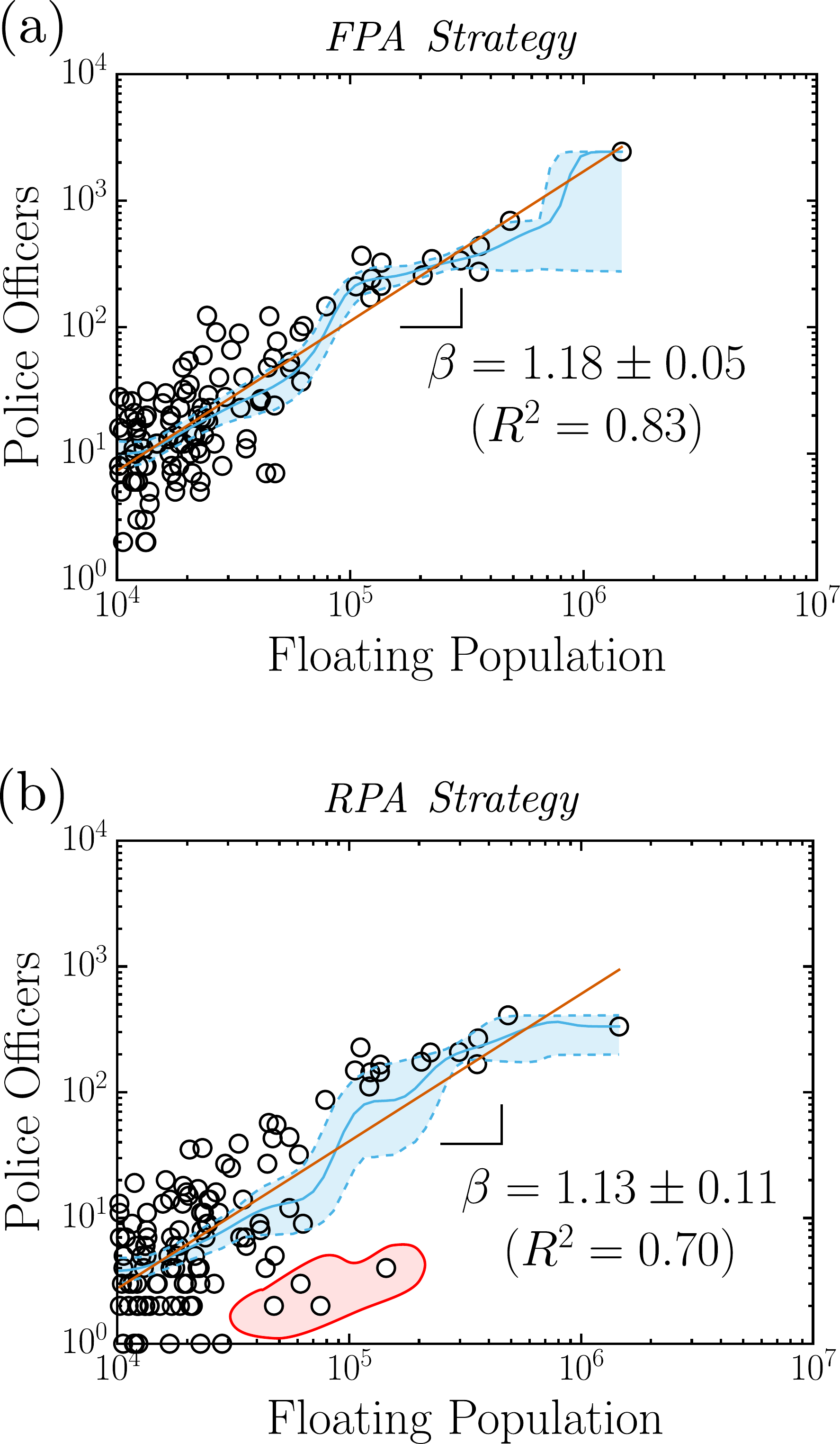}
\caption{{\bf Correlations between police officers and floating population in
\textit{RPA and FPA Strategy}.} (a) and (b), respectively, illustrates the
correlations achieved for \textit{RPA and FPA Strategy}. The $x$-axis represents
the floating population and $y$-axis the number of police officers allocated.
The red lines represent the simple linear regressions applied to the data, the
blue continuous lines represent the Nadaraya-Watson method
\cite{nadaraya1964,watson1964} and the blue dashed lines delimit the 95\%
confidence interval (CI) estimated by bootstrap}. 
\label{fig6} 
\end{figure}

Such difference quantifies the inefficacy of the {\it RPA Strategy}. While the
allocation produced from the {\it FPA strategy} is strongly correlated with the
flow of people, the {\it RPA strategy} fails to capture the scale found by
Caminha {\it et al.} \cite{caminha2017}. Remember that their studies found a
superlinear relationship between property crimes and floating population with
exponent $\beta = 1.15 \pm 0.4$. Figure \ref{fig6} shows the correlation between
the number of police resources allocated and floating population following the
two different strategies. In (a) it is shown the correlation between the
resources allocated and floating population following the FPA strategy. There is
a clear superlinear relation with an exponent of $\beta = 1.18 \pm 0.05$ and a
strong coefficient of determination \cite{rawlings2001, montgomery2015} ($R^2 =
0.83$), On the other hand, in (b), although a superlinear relation appears, the
determination coefficient ($R^2 = 0.70$) as well as the standard error of this
\cite{rawlings2001, montgomery2015} ($\pm 0.11$) reveals that the {\it RPA
Strategy} is not the more adequate to the city of Fortaleza. Another important
feature that indicates the inappropriateness of the {\it RPA strategy} is also
observed in Figure \ref{fig6}, specifically the analysis of the dispersion of the dots
(clusters). In (b), we can see four clusters of floating population with
considerable activity (flow of people) with few police resources. This happens
because the boundaries between neighborhoods sometimes divide the floating
clusters what makes difficult a precise allocation of resource in that region.
In general, although the {\it RPA strategy} suggests the distribution of
resources in a way that follows a Power Law, there is an imprecision because
this strategy aims at capturing the influence of floating population indirectly
via the incidence of crime. That is to say, as crime occurs due to the presence
of people, looking at crime is a way to consider the floating population. This
is not however the best approach because it fails to capture the potential of
occurrence of crime in a disproportional way caused by the existence of clusters
of floating population. When the {\it FPA Strategy} is applied the cause (flow
of people in a region) and the amount of crime are considered to determine the
amount of resources to be allotted. Doing so, it is possible to statistically
approximate (in terms of exponent and standard error) the superlinear relation
as suggested by Caminha {\it et al.} \cite{caminha2017}.

\section{Conclusion}
This paper presented a study that investigates new ways of allocating police
resources within the urban space. Differently to conventional allocation
policies, which allocate resources through the city using administrative units ,
an allocation strategy was presented which distributes police by clusters of
floating population, which have already been proved to be much more precise in
explaining the behavior of crimes against property in a city \cite{caminha2017}.
This precision is due to the fact that the borders of population flux often go
beyond the boundaries of the administrative divisions and clustering algorithms
identify the ''islands'' formed by those clusters that are naturally strategic
regions in combating crime.

Our study reveals that allocation of police resources into clusters of floating
population leads the distribution of resources in a way significantly different
from strategies that allocates resources having per basis the administrative
regions. More specifically, we show that the allocation having as basis the
clusters of floating population tends to be more adequate for fighting crime
against properties because the distribution of police resources will naturally
follow a Power Law, what is desirable since it is expected that crime grows
disproportionally in areas with high density of floating population.

The aspects discussed here open new lines of further investigations. In
particular, it is important to notice that the work by Caminha et al. has also
shown that for certain types of crimes (e.g. peace disturbance) the superlinear
relationship is only captured having as basis administrative areas that account
for features of resident population rather than clusters of floating population.
This indicates that it is necessary to think in a hybrid strategy in which
different polices and different divisions of the urban space need to be
taken into consideration for each type of crime.

\ifCLASSOPTIONcaptionsoff
  \newpage
\fi


\end{document}